\documentstyle[preprint,aps,axodraw]{revtex}
\tightenlines

\newcommand{\reseteqnum}{\setcounter{equation}{0}}
\newcommand{\nn}{\nonumber}
\newcommand{\eqn}[1]{(\ref{#1})}
\newcommand{\ovl}[1]{\overline{#1}}
\newcommand{\wt}[1]{\widetilde{#1}}
\newcommand{\p}{\partial}
\newcommand{\pslash}{p\kern-1ex /}
\newcommand{\lslash}{l\kern-1ex /}
\newcommand{\Dslash}{{\cal D}\kern-1.5ex /}
\newcommand{\bpsi}{\overline{\psi}}
\newcommand{\vev}[1]{\left\langle #1 \right\rangle}
\def\setcaption#1{\def\@captype{#1}}

\begin{document}


\title{
\vspace{-3.0cm}
\begin{flushright}  
{\normalsize UTHEP-393}\\
\end{flushright}
Perturbative Renormalization Factors of Quark Bilinear Operators 
for Domain-wall QCD}

\author{$^1$Sinya Aoki, $^1$Taku Izubuchi, $^{2}$Yoshinobu Kuramashi
\thanks{On leave from Institute of Particle and Nuclear Studies,
High Energy Accelerator Research Organization(KEK),
Tsukuba, Ibaraki 305-0801, Japan}
 and $^1$Yusuke Taniguchi}

\address{$^1$Institute of Physics, University of Tsukuba, 
Tsukuba, Ibaraki 305-8571, Japan \\
$^2$Department of Physics, Washington University, 
St. Louis, Missouri 63130, USA \\
}

\date{\today}

\maketitle

\begin{abstract}
We calculate one-loop renormalization factors of bilinear operators 
made of physical quark fields for domain-wall QCD.
We find that finite parts of such renormalization factors
have reasonable values at 1-loop
except an overlap factor between the physical quark field and the zero mode 
in the theory. We point out that the 1-loop estimate of
overall renormalization factors becomes unreliable at the coupling
where numerical simulations are currently performed,
due to the presence of this overlap factor.
We show that this problem disappears if the mean-field improved
perturbation theory is employed for renormalization factors.
\end{abstract}


\narrowtext

\section{Introduction}

The lack of chirally invariant fermion formulations is one of the most
uncomfortable points theoretically and practically in lattice QCD.
For example,
in the Wilson fermion formulation, which is popularly used in numerical 
simulations,
the chiral limit can be realized only by
the fine tuning of bare mass parameter, which compensates
the additive quantum correction to the quark mass.

Recently the domain-wall fermion formulation\cite{Shamir93,Shamir95},
which was originally proposed for lattice chiral gauge theories\cite{Kaplan}, 
has been employed in lattice QCD simulations\cite{Blum-Soni}
and has shown its superiority over other formulations:
there seems no need of the fine tuning to realize the chiral limit
while there is no restriction to the number of flavors. 
In particular massless mode which presents at the tree level seems stable
against the quantum correction. Since this property is not a trivial one,
two of us have performed the perturbative calculation for the domain-wall
quark propagator and have shown that the massless mode is indeed stable against
the 1-loop correction\cite{Aoki-Taniguchi}. 
The wave function renormalization factor for the 
massless mode has also been evaluated there.

In this paper we extend our previous perturbative calculation to an
evaluation of 
renormalization factors for quark bilinear operators at 1-loop 
in domain-wall QCD. These renormalization factors are needed to convert
matrix elements such as meson decay constants measured in lattice simulations 
to those in $\ovl{\rm MS}$ scheme, so that the values predicted on the
lattice can be compared with the corresponding experimental values.
In particular, in order to discuss the scaling behavior of such matrix
elements, the correction by the renormalization factors becomes
important, since 1-loop corrections are in general non-negligible at
the gauge coupling used in simulations. Moreover since
1-loop corrections decrease only as the inverse of the logarithm of the 
lattice spacing, not a power of it, they can not be removed through
the usual continuum extrapolation like 
linear or quadratic in the lattice spacing.
Therefore the corrections by the renormalization factors to 
such matrix elements must be included for precise measurements
in numerical simulations with the domain-wall QCD.

We start from the Green's function with quark bilinear operators and
physical quark fields.
We show that the divergence in the Green's function can be renormalized
into the quark propagator and bilinear operators.
There appears no mixing between these operators and operators 
made of heavy unphysical fermions.
The renormalization factor consists of two parts, one is the usual
renormalization factor which has similar value as that of Wilson fermion,
another is the factor of overlap between field on each two boundary wall 
and the zero mode.
We notice that this overlap factor is renormalized additively and tends to
be large.
We present renormalization factors with mean-field improvement to suppress
this large correction.

This paper is organized as follows.
In Sec.~II we introduce the action of domain-wall QCD 
together with some notations used in this paper
and explicitly give the Feynman rules relevant to our calculations.
In Sec.~III we evaluate the self-energy of the physical fermion, from which
we extract the wave function renormalization factor of the physical fermion 
and the multiplicative mass renormalization factor. Although results
for both wave function\cite{Aoki-Taniguchi} and 
mass\cite{Wingate} renormalization factors
have already been reported, we give them here in order to demonstrate
our method of calculating renormalization factors in domain-wall QCD.
In particular we pay attention to the extra factor in the wave function 
renormalization, which arises as the overlap between the physical fermion
field and the zero mode. 
We then calculate the renormalization factors for quark bilinear operators in 
Sec.~IV.
Numerical results of one-loop coefficients for various
quantities are given in Sec.~V as a function of $M$
in both cases with and without  the mean-field improvement.
Our conclusion is given in Sec.~VI.

In this paper we set the lattice spacing $a = 1$ and take $SU(N_c)$ gauge
group with the gauge coupling $g$ and the second Casimir $C_F = 
\displaystyle \frac{N_c^2-1}{2N_c}$.

\section{Action and Feynman rules}

The domain-wall fermion is a 4+1 dimensional Wilson fermion with
a ``mass term''  which depends on the coordinate in the extra dimension.
For the explicit form of the ``mass term'' 
we adopt the Shamir's one\cite{Shamir93} in this paper and the quark
action with the current quark mass $m$ becomes
\begin{eqnarray}
S_{\rm DW} &=&
\sum_{n} \sum_{s=1}^N \Biggl[ \frac{1}{2} \sum_\mu
\left( \bpsi(n)_s (-r+\gamma_\mu) U_\mu(n) \psi(n+\mu)_s
+ \bpsi(n)_s (-r-\gamma_\mu) U_\mu^\dagger(n-\mu) \psi(n-\mu)_s \right)
\nn\\&&
+ \frac{1}{2}
\left( \bpsi(n)_s (1+\gamma_5) \psi(n)_{s+1}
+ \bpsi(n)_s (1-\gamma_5) \psi(n)_{s-1} \right)
+ (M-1+4r) \bpsi(n)_s \psi(n)_s \Biggr]
\nn\\&+&
 m \sum_n \left( \bpsi(n)_{N} P_{+} \psi(n)_{1}
+ \bpsi(n)_{1} P_{-} \psi(n)_{N} \right),
\end{eqnarray}
where $n$ is a 4 dimensional space-time coordinate and $s$ is an extra 
fifth dimensional index,
the Dirac ``mass'' $M$ is a parameter of the theory
which we set $0 \le M \le 2$ to realize the massless fermion at tree level,
$m$ is a physical quark mass,
and the Wilson parameter is set to $r=-1$.
Our $\gamma$ matrix convention is as follows:
\begin{eqnarray}
&&
\gamma_i
= \pmatrix{
  0        & -i \sigma^i \cr
i \sigma^i & 0           \cr
},
\quad
\gamma_4
= \pmatrix{
0 & 1 \cr
1 & 0 \cr
},
\quad
\gamma_5 \equiv \gamma_1 \gamma_2 \gamma_3 \gamma_4
=\pmatrix{
1 &  0 \cr
0 & -1 \cr
} , \\
&&
P_\pm = \frac{1 \pm \gamma_5}{2}, \\
&&
\sigma_{\mu \nu} = \frac{1}{2} \left[ \gamma_\mu , \gamma_\nu \right] . 
\end{eqnarray}
The gauge part of the domain-wall fermion is a standard 4 dimensional
one plaquette action and we have no gauge interaction along the fifth
dimension.
We can interpret this fifth dimensional degrees of freedom as a flavor
\cite{NN}.
It is important to notice that we have boundaries for the flavor space;
$1 \le s \le N$.

The remarkable property of the domain-wall fermion is that there exists
a massless fermion mode in the $N\to\infty$ limit at small momentum
at $m=0$.
This massless fermion stays near the boundaries of
the flavor space with the left and the right mode on the opposite side.
At tree level the massless mode $\chi_0$ is given explicitly in
zero momentum limit as
\begin{eqnarray}
\chi_0 = \sqrt{1-w_0^2}
\left( P_+ w_0^{s-1} \psi_s + P_- w_0^{N-s} \psi_s \right),
\end{eqnarray}
where $w_0 = 1-M$.
Although the zero mode is stable against quantum
correction\cite{Aoki-Taniguchi}, the damping factor $w_0$ is modified 
because the Dirac mass $M$ is renormalized additively.
In numerical simulations
it is more convenient to use the interpolating ``physical'' quark field
defined by the boundary fermions
\begin{eqnarray}
q(n) = P_+ \psi(n)_1 + P_- \psi(n)_N,
\nn \\
\ovl{q}(n) = \bpsi(n)_N P_+ + \bpsi(n)_1 P_-.
\label{eq:quark}
\end{eqnarray}
We use the QCD operators constructed from this quark fields,
since this field has been actually used in the previous simulations.
Moreover, as seen later, we find that the renormalization becomes
simpler for the operators in terms of the physical quark field.

Weak-coupling perturbation theory is developed by writing
\begin{equation}
U_{n,\mu}=\exp (igaA_\mu(n+\frac{1}{2}\hat\mu)) .
\end{equation}
Since our gauge part is same as that of the usual Wilson plaquette action,
the gluon propagator can be written as
\begin{eqnarray}
G_{\mu \nu}^{ab} (p)
=\frac{1}{4\sin^2 p/2}
\left[\delta_{\mu \nu}
- (1-\alpha) \frac{4 \sin {p}_\mu/2 \sin {p}_\nu/2}{4 \sin^2 p/2}
\right]
 \delta_{ab},
\end{eqnarray}
where $\sin^2 p/2 = \sum_\mu \sin^2 p_\mu/2$ and
we set $\alpha = 1$ in our calculation.
Quark-gluon vertices are also identical to those in 
the $N$ flavor Wilson fermion. We employ the following two of those
relevant for the one loop calculation:
\begin{eqnarray}
V_{1\mu}^a (k,p)_{st}
&=& -i g T^a \{ \gamma_\mu \cos \frac{a}{2}(-k_\mu + p_\mu)
  -i r \sin \frac{a}{2}(-k_\mu + p_\mu) \} \delta_{st},
\\
V_{2\mu\nu}^{ab} (k,p)_{st}
&=& \frac{a}{2} g^2 \frac{1}{2} \{T^{a}, T^{b}\}
\{ i \gamma_\mu \sin \frac{a}{2}(-k_\mu + p_\mu)
-r \cos \frac{a}{2} (-k_\mu + p_\mu) \}\delta_{\mu\nu} \delta_{st}.
\end{eqnarray}
Our momentum assignments for the vertices are given in Fig.~\ref{fig:vertex}.

The fermion propagator is given by inverting the domain-wall Dirac 
operator, whose formula in momentum space is
\begin{eqnarray}
\Dslash(p)_{st} = \sum_\mu i \gamma_\mu \sin p_\mu \delta_{st}
+ \left(W^+ (p) + m M^+\right)_{s,t} P_+
+ \left(W^- (p) + m M~-\right)_{s,t} P_-,
\label{eqn:Dirac-Op}
\end{eqnarray}
where the mass matrix is
\begin{eqnarray}
&&
W^{+} (p)_{s,t} =
\pmatrix{
-W(p) & 1     &        &       \cr
      & -W(p) & \ddots &       \cr
      &       & \ddots & 1     \cr
      &       &        & -W(p) \cr
},
\label{eqn:mass-matrix-p}
\\&&
W^{-} (p)_{s,t} =
\pmatrix{
-W(p) &        &        &       \cr
1     & -W(p)  &        &       \cr
      & \ddots & \ddots &       \cr
      &        & 1      & -W(p) \cr
},
\label{eqn:mass-matrix-m}
\\&&
M^+ = \pmatrix{
  &  &  \cr
  &  &  \cr
1 &  &  \cr}
,\quad
M^- = \pmatrix{
  &  & 1\cr
  &  &  \cr
  &  &  \cr},
\\&&
W(p) = 1-M -r \sum_\mu (1-\cos p_\mu).
\end{eqnarray}
In numerical simulations one should take $N$ large enough in order
to have a massless quark in the spectrum at $m=0$.
This means that we can effectively neglect $e^{-\alpha N}$ terms with
positive $\alpha$. 
Therefore in our 1-loop calculation we will take $N\to\infty$ limit 
to avoid complications arising from the finite $N$.
Although the $N\to\infty$ limit should be taken after the momentum integral
in principle,
it is easy to see that the limit can be taken before the integral.
We may take the form of  free fermion propagator 
in the $N\to\infty$ limit from the beginning.
(See \cite{Aoki-Taniguchi} for the derivation of the finite $N$ fermion
propagator.)
Inverting \eqn{eqn:Dirac-Op} the fermion propagator 
for $N\to\infty$ becomes
\begin{eqnarray}
S_F(p)_{st} &\equiv& \left(\Dslash(p)^{-1}\right)_{st}
\nn\\&=&
\left( -i\gamma_\mu \sin p_\mu + W^- + m M^+ \right)_{su} G_R(u,t) P_+
\nn\\&+&
\left( -i\gamma_\mu \sin p_\mu + W^+ + m M^- \right)_{su} G_L(u,t) P_-,
\end{eqnarray}
where sum over the same index is taken implicitly.
$G_{R/L}$ is given by 
\begin{eqnarray}
G_{R} (s, t) 
&=&
\frac{A}{F}
\Bigl[
-(1-m^2) \left(1-W e^{-\alpha}\right) e^{\alpha (-2N+s+t)}
-(1-m^2) \left(1-W e^\alpha\right) e^{-\alpha (s+t)}
\nn\\&&
-2W \sinh (\alpha) m
 \left( e^{\alpha (-N+s-t)} + e^{\alpha (-N-s+t)} \right)
\Bigr]
+ A e^{-\alpha |s-t|},
\\
G_{L} (s, t) 
&=&
\frac{A}{F}
\Bigl[
-(1-m^2) \left(1-We^{\alpha}\right) e^{\alpha (-2N+s+t-2)}
-(1-m^2) \left(1-We^{-\alpha}\right) e^{\alpha (-s-t+2)}
\nn\\&&
-2W \sinh (\alpha) m
 \left( e^{\alpha (-N+s-t)} +  e^{\alpha (-N-s+t)} \right)
\Bigr]
+ A e^{-\alpha |s-t|},
\\
\cosh (\alpha) &=& \frac{1+W^2+\sum_\mu \sin^2 p_\mu}{2W},
\\
A&=& \frac{1}{2W \sinh (\alpha)},
\\
F &=& 1-e^{\alpha} W-m^2 \left(1-W e^{-\alpha}\right).
\end{eqnarray}
Note that the argument $p$ of factors $\alpha$ and $W$ is suppressed
in the above formula.

Besides the fermion propagator given above,
the propagator for the physical quark field, defined in \eqn{eq:quark},
is also used in this paper. The explicit form is rather simple and is given by
\begin{eqnarray}
S_q(p) \equiv \vev{q(-p) \ovl{q}(p)} = 
 \frac{-i\gamma_\mu \sin p_\mu + \left(1-W e^{-\alpha}\right) m}
{-\left(1-e^{\alpha}W\right) + m^2 (1-W e^{-\alpha})}.
\label{eqn:phys-prop}
\end{eqnarray}

\reseteqnum
\section{Renormalization for quark mass and wave function}

In this section we calculate the self energy for the physical quark field,
 from which we derive renormalization factors for mass and wave function.
The physical quark propagator \eqn{eqn:phys-prop} at tree level becomes 
\begin{eqnarray}
S_q(p) &=&
\frac{(1-w_0^2)}{i\pslash + (1-w_0^2) m}
\end{eqnarray}
in the continuum limit.
In the following we will see that the full quark propagator with 1-loop
corrections takes the following form:
\begin{eqnarray}
S_q(p)_{\rm full} &=&
\frac{(1-w_0^2)Z_w Z_2}{i\pslash + (1-w_0^2)Z_w Z_m^{-1} m},
\end{eqnarray}
where $Z_2$ and $Z_m$ are the quark wave function and the mass renormalization
factors respectively. In Ref.~\cite{Aoki-Taniguchi} $Z_2$ is calculated,
while $Z_m$ is given in Ref.~\cite{Wingate}.
$Z_w$, which is dropped in Ref.~\cite{Aoki-Taniguchi}, 
is a renormalization factor for the overall factor $(1-w_0^2)$.

One loop corrections to the quark two point function
$\vev{q(-p) \ovl{q}(p)}$ are given by two diagrams in
 Fig.~\ref{fig:propagator}.
In the diagrams interaction vertices contain fermion field
of a general flavor index, while
physical quark field appears on the external lines.
Tree-level propagators for the external lines are given by
\begin{eqnarray}
&&
\vev{q(-p) \bpsi(p,s)} 
=
\frac{1}{F}
\left( i\gamma_\mu \sin p_\mu - m \left(1 -W e^{-\alpha} \right)
\right)
\left( e^{-\alpha (N-s)} P_+ + e^{-\alpha (s-1)} P_- \right)
\nn\\&&\qquad
+\frac{1}{F} \Bigl[
m \left(i\gamma_\mu \sin p_\mu  -m \left(1-W e^{-\alpha}\right)\right)
- F \Bigr] e^{-\alpha}
\left( e^{-\alpha (s-1)} P_+ + e^{-\alpha (N-s)} P_- \right),
\\&&
\vev{\psi(-p,s) \ovl{q}(p)} 
=
\frac{1}{F}
\left( e^{-\alpha (N-s)} P_- + e^{-\alpha (s-1)} P_+ \right)
\left( i\gamma_\mu \sin p_\mu - m \left(1 - W e^{-\alpha} \right)
\right)
\nn\\&&\qquad
+\frac{1}{F}
\left( e^{-\alpha (s-1)} P_- + e^{-\alpha (N-s)} P_+ \right) e^{-\alpha}
\Bigl[
m \left(i\gamma_\mu \sin p_\mu  -m\left(1- We^{-\alpha}\right) \right)
- F \Bigr].
\end{eqnarray}
In the one loop calculation we take external momenta and quark masses
much smaller than the lattice cut-off, so that
we can expand the propagator corresponding to external lines
in terms of external momenta and masses.
In  diagrams in Fig.~\ref{fig:propagator} it is enough for the
renormalization to expand the external propagator to the next to leading
order,
\begin{eqnarray}
&&
\vev{q(-p) \bpsi(p,s)}
\nn\\&&\quad
\to
\frac{1-w_0^2}{i\pslash +(1-w_0^2)m}
\Biggl[
\left( w_0^{(N-s)} P_+ + w_0^{(s-1)} P_- \right)
-\frac{w_0}{1-w_0^2} i\pslash 
\left( w_0^{(s-1)} P_+ + w_0^{(N-s)} P_- \right) \Biggr],
\label{eqn:ext-line1}
\\&&
\vev{\psi(-p,s) \ovl{q}(p)} 
\nn\\&&\quad
\to
\Biggl[
 \left( w_0^{(N-s)} P_- + w_0^{(s-1)} P_+ \right)
-\left( w_0^{(s-1)} P_- + w_0^{(N-s)} P_+ \right)
\frac{w_0}{1-w_0^2} i\pslash \Biggr]
\frac{1-w_0^2}{i\pslash + (1-w_0^2) m}.
\label{eqn:ext-line2}
\end{eqnarray}
Note that both terms at the leading and the next to leading orders in the
external propagator give the leading order contribution in these diagrams 
since interaction vertices have
the form $a + b p_\mu$ due to the Wilson term.

The external propagator has the characteristic form in the continuum limit
that it consists of the quark propagator times some dumping factor.
One loop correction to the quark propagator becomes
\begin{eqnarray}
\vev{q(-p) \ovl{q}(p)}_1
=
\frac{1-w_0^2}{i\pslash +(1-w_0^2)m}
\Sigma_q(p,m)
\frac{1-w_0^2}{i\pslash +(1-w_0^2)m},
\label{eqn:one-loop}
\end{eqnarray}
which is easily renormalized into the quark propagator.
Here $\Sigma_q(p,m)$ is given by multiplying the fermion self energy
$\Sigma(p,m)_{st}$ by the factors in the external propagator,
\begin{eqnarray}
\Sigma_q(p,m) &=&
\left[
\left( w_0^{(N-s)} P_+ + w_0^{(s-1)} P_- \right)
-\frac{w_0}{1-w_0^2} i\pslash 
\left( w_0^{(s-1)} P_+ + w_0^{(N-s)} P_- \right) \right]
\Sigma(p,m)_{st}
\nn\\&&\times
\left[
 \left( w_0^{(N-t)} P_- + w_0^{(t-1)} P_+ \right)
-\left( w_0^{(t-1)} P_- + w_0^{(N-t)} P_+ \right)
\frac{w_0}{1-w_0^2} i\pslash \right].
\label{eqn:sigmaq1}
\end{eqnarray}
Calculating the quark self-energy $\Sigma(p,m)_{st}$ to one-loop order 
and making an expansion of the form 
\begin{equation}
\Sigma(p,m)_{st} = \Sigma(0)_{st}
+\frac{\partial\Sigma(0)_{st}}{\partial p_\mu} p_\mu
+\frac{\partial\Sigma(0)_{st}}{\partial m} m
+O(p^2,m^2,pm),
\end{equation}
we find 
\begin{eqnarray}
\Sigma(p,m)_{st} &=&
-g^2 C_F \left[
i \pslash \left(I^+ P_+ + I^- P_-\right)
+ W^+_1 P_+ + W^-_1 P_-
+ m \left( M^+_1 P_+ + M^-_1 P_- \right)
\right]_{st},
\nn\\
\end{eqnarray}
where $I^\pm$ and $W_1^\pm$ are already calculated in our previous paper
\cite{Aoki-Taniguchi} for the massless fermion,
\begin{eqnarray}
I^{+/-}_{st}
&=& -\frac{T}{2} \delta_{st}
+\frac{1}{16\pi^2} (C_{+/-})_{s,t}
\left( \ln (\pi^2) -1 - \ln \lambda^2 \right)
\nn\\&+&
\int \frac{d^4 l}{(2\pi)^4}
\frac{1}{4 \sin^2 l/2}
\Biggl[
\frac{1}{8}\sum_\mu\left(
\cos l_\mu ( W^-G_R + W^+G_L)(s,t) + \sin^2l_\mu (G_L+G_R)(s,t) 
\right) \nn\\
&+& \sum_\mu \frac{\sin^2 l_\mu}{4(4\sin^2 l/2)^2}
\left(
( W^-G_R + W^+G_L)(s,t) + 2(\sum_\nu \cos^2 \frac{l_\nu}{2} -
2\cos^2 \frac{l_\mu}{2})G_{L/R}(s,t) \right. \nn \\
&+& \left. \sum_\nu \sin^2 \frac{l_\nu}{2} G_{R/L}(s,t)
\right) \Biggr]
- (C_{+/-})_{s,t} \int \frac{d^4 l}{(2\pi)^4}
\frac{1}{(l^2)^2} \theta (\pi^2-l^2) ,
\label{eqn:finite-p}
\\
(W^{+/-}_1)_{s,t}
&=&
-2T \delta_{st}
+\int \frac{d^4 l}{(2\pi)^4}
\frac{1}{4\sin^2 l/2}
\sum_\mu
\Biggl[
\cos^2 \frac{l_\mu}{2} (W^{+/-}G_{L/R})(s,t)
\nn\\&-&
\sin^2 \frac{l_\mu}{2}(W^{-/+}G_{R/L})(s,t)
+\frac{1}{2}\sin^2 l_\mu ( G_L+G_R)(s,t)
\Biggr] .
\label{eqn:linear-div-p}
\end{eqnarray}
Here $T$ is the tadpole loop integral
\begin{eqnarray}
T = \int_{-\pi}^\pi \frac{d^4 l}{(2 \pi)^4}
\frac{1}{4 \sin^2 \frac{l}{2}} = 0.15493.
\end{eqnarray}
Note that the notation used in this paper is slightly different from
that in Ref.~\cite{Aoki-Taniguchi}.
In particular 
the infrared divergence in the loop integral is 
regularized by a gluon mass $\lambda$ introduced in the gluon propagator
in this paper,
while the external fermion momentum is kept non-zero in 
Ref.~\cite{Aoki-Taniguchi} to avoid such an infrared divergence.
The calculation becomes simpler in the former.

It is also straightforward to calculate one loop correction to mass term
$M_1^\pm$,
\begin{eqnarray}
(M^{+/-}_1)_{s,t}
&=&
\frac{4}{16\pi^2} (B_{+/-})_{s,t}
\left( \ln (\pi^2) - \ln \lambda^2 \right)
\nn\\&+&
\int \frac{d^4 l}{(2\pi)^4}
\frac{1}{4\sin^2 l/2}
\sum_\mu
\Biggl[
 \cos^2 \frac{l_\mu}{2} \left( W^{+/-}\frac{\p G_{L/R}}{\p m}
 + M^{+/-}G_{L/R} \right)(s,t)
\nn\\&-&
\sin^2 \frac{l_\mu}{2} \left( W^{-/+}\frac{\p G_{R/L}}{\p m}
 + M^{-/+}G_{R/L} \right)(s,t)
+\frac{1}{2}\sin^2 l_\mu
\left( \frac{\p G_L}{\p m}+\frac{\p G_R}{\p m}\right)(s,t)
\Biggr]
\nn\\&-&
 (B_{+/-})_{s,t} 4 \int \frac{d^4 l}{(2\pi)^4}
\frac{1}{(l^2)^2} \theta (\pi^2-l^2) ,
\end{eqnarray}
where $C_{+/-}$ and $B_{+/-}$ are defined by
\begin{eqnarray}
&&
(C_+)_{st}=(1-w_0^2) w_0^{s+t-2},\quad
(C_-)_{st}=(1-w_0^2) w_0^{2N-s-t},
\\&&
(B_+)_{st}=(1-w_0^2)^2 w_0^{N-s+t-1},\quad
(B_-)_{st}=(1-w_0^2)^2 w_0^{s-1+N-t}.
\end{eqnarray}

Substituting these into \eqn{eqn:sigmaq1} and neglecting 
higher order terms of ${\cal O}(p^2,m^2,pm)$ we obtain
\begin{eqnarray}
\Sigma_q(p,m) &=&
-\frac{1}{1-w_0^2} i\pslash
\frac{g^2 C_F}{16\pi^2}
\Biggl[
\left(-\log \lambda^2 a^2 - \Sigma_1 \right)
+ \frac{2w_0}{1-w_0^2} \Sigma_3
\Biggr]
\nn\\&-&
m \frac{g^2 C_F}{16\pi^2}
\left(-4\log \lambda^2 a^2 - \Sigma_2 \right),
\label{eqn:sigmaq2}
\end{eqnarray}
where
\begin{eqnarray}
\Sigma_1 &=& -\log \pi^2 +1 +(\frac{T}{2} -I^d)\cdot 16\pi^2 ,
\\
\Sigma_2 &=& -4\log \pi^2 -\sigma_m\cdot 16\pi^2 ,
\\
\Sigma_3 &=& (2T -\wt{\omega})\cdot 16\pi^2 ,
\\
I^d & = & \int\frac{d^4l}{(2\pi)^4}\left\{
\frac{1}{32\sin^2 l/2} \sum_\mu \left[  \sin^2l_\mu ( \wt{G}_R+\wt{G}_L)
+ 2 \cos l_\mu (w_0-W)\wt{G}_R \right] \right. \nn \\
&+& \sum_\mu \frac{\sin^2 l_\mu}{2(4\sin^2 l/2)^2} 
\left[(w_0-W)\wt{G}_R+(\sum_\nu \cos^2 l_\nu/2 - 2\cos^2l_\mu/2 )
\wt{G}_L+\sum_\nu (\sin^2l_\nu/2) \wt{G}_R \right] \nn \\
&-&\left. \frac{1}{(l^2)^2}\theta(\pi^2-l^2)\right\} ,
\\
\wt{\omega} &=&
\int \frac{d^4 l}{(2\pi)^4}
\frac{1}{4\sin^2 l/2}
\sum_\mu
\Biggl[
 \cos^2 \frac{l_\mu}{2} (w_0-W) \wt{G}_R
-\sin^2 \frac{l_\mu}{2} (w_0-W) \wt{G}_R
\nn \\
&+&
\frac{1}{2}\sin^2 l_\mu (\wt{G}_L+\wt{G}_R)
\Biggr],
\\
\sigma_m &=&
\int \frac{d^4 l}{(2\pi)^4}
\frac{1}{4\sin^2 l/2}
\frac{1}{\left(1-w_0 e^{-\alpha}\right)^2}
\sum_\mu
\Biggl[
-\cos^2 \frac{l_\mu}{2}
\frac{1-e^{-\alpha} W}{1-e^{\alpha} W}
\nn\\&+&
r^2 \sin^2 \frac{l_\mu}{2} e^{-2\alpha}
+r\sin^2 l_\mu \frac{e^{-\alpha}}{1-e^\alpha W}
\Biggr]
- 4 \int \frac{d^4 l}{(2\pi)^4} \frac{1}{(l^2)^2} \theta(\pi^2-l^2).
\end{eqnarray}
The propagators $\wt{G}_L$ and $\wt{G}_R$ are defined by
\begin{eqnarray}
\wt{G}_L &=&
\frac{1-w_0^2}{2W\sinh\alpha}
\left[
\frac{\sinh\alpha_0 -\sinh\alpha}{2w_0\sinh\alpha_0 (\cosh\alpha_0
-\cosh\alpha)}
-\frac{e^\alpha-W}{e^{-\alpha}-W}\frac{1}{(e^\alpha-w_0)^2}
\right] ,
\\
\wt{G}_R &=&
\frac{1-w_0^2}{2W\sinh\alpha}
\left[
\frac{\sinh\alpha_0 -\sinh\alpha}{2w_0\sinh\alpha_0 (\cosh\alpha_0
-\cosh\alpha)}
-\frac{1}{(e^\alpha-w_0)^2}
\right] 
\end{eqnarray}
with $e^{-\alpha_0}=w_0$.
It should be noted that there is no additive mass correction 
in Eq.~\eqn{eqn:sigmaq2}.
Although the fermion self energy $\Sigma(p,m)_{st}$ has an additive
mass correction, it vanishes in $\Sigma_q$ because of the relation,
\begin{eqnarray}
\left( w_0^{N-s} P_+ + w_0^{s-1} P_- \right)
\left( W_1^+ P_+ + W_1^- P_- \right)_{st}
\left( w_0^{N-t} P_- + w_0^{t-1} P_+ \right)
= 0,
\end{eqnarray}
which has been shown in the previous paper\cite{Aoki-Taniguchi}
\footnote{
At finite $N$ we need a few extra terms for the renormalization such as
an additive mass counter term.
However these terms always contain the factor $e^{-\alpha N}$ and
can be suppressed in the large $N$ limit before taking the continuum limit.
}
.

The full quark propagator at 1-loop level now becomes
\begin{eqnarray}
S_q(p)_{\rm full} &=& \frac{1-w_0^2}{i\pslash + (1-w_0^2)m} \left[
1 + \Sigma_q (p, m)\frac{1-w_0^2}{i\pslash + (1-w_0^2)m}
\right] \nn \\
&=& \frac{1-w_0^2}{i\pslash +(1-w_0^2)m-(1-w_0^2)\Sigma_q(p,m)} + O(g^4) ,
\end{eqnarray}
and finally we obtain
\begin{eqnarray}
S_q(p)_{\rm full} 
&=& \frac{(1-w_0^2) Z_w Z_2}{i\pslash + (1-w_0^2) Z_w Z_m^{-1} m}
\end{eqnarray}
with
\begin{eqnarray}
Z_2 &=& 
1+\frac{g^2 C_F}{16\pi^2} \left(\log \lambda^2 a^2 +\Sigma_1 \right),
\\
Z_w &=& 
1-\frac{2w_0}{1-w_0^2} \frac{g^2 C_F}{16\pi^2} \Sigma_3
=1+\frac{g^2 C_F}{16\pi^2} z_w ,
\label{eqn:zw}
\\
Z_m^{-1} &=&
1 + \frac{g^2 C_F}{16\pi^2} \left(
-3\log \lambda^2 a^2 - \Sigma_2 + \Sigma_1 \right).
\end{eqnarray}
 From this relation we can see that $\Sigma_3$ is an additive
renormalization to $w_0$,
\begin{eqnarray}
\left(1-w_0^2\right) Z_w = 1-
\left(w_0+\frac{g^2 C_F}{16\pi^2} \Sigma_3\right)^2.
\end{eqnarray}

If we employ the mean-field (tadpole) improvement\cite{MF}, we can write
\begin{eqnarray}
S_q(p) &=& \frac{(1-(w^{\rm MF}_0)^2) Z^{\rm MF}_w u^{-1} Z^{\rm MF}_2}
{i\pslash + (1-(w^{\rm MF}_0)^2)
Z^{\rm MF}_w (u Z^{\rm MF}_m)^{-1} m},
\end{eqnarray}
where $u = 1-g^2 C_F T/2$, $w^{\rm MF}_0=w_0-4(u-1)$, 
\begin{eqnarray}
Z^{\rm MF}_2 & = &  1+\frac{g^2 C_F}{16\pi^2} 
\left(\log \lambda^2 a^2 +\Sigma_1 -16\pi^2 \frac{T}{2} \right) , \\
Z^{\rm MF}_w &=& 1-\frac{2w_0}{1-w_0^2} \frac{g^2 C_F}{16\pi^2} 
\left(\Sigma_3 - 16\pi^2\cdot 2 T \right) 
\equiv 1+\frac{g^2 C_F}{16\pi^2} z^{\rm MF}_w , \\
(Z^{\rm MF}_m)^{-1} &=&
1 + \frac{g^2 C_F}{16\pi^2} \left(
-3\log \lambda^2 a^2 - \Sigma_2 + \Sigma_1 -16\pi^2 \frac{T}{2} \right).
\end{eqnarray}
Of course equalities that $u Z_2 = Z^{\rm MF}_2$, 
$(1-w_0^2)Z_w = (1-(w^{\rm MF}_0)^2)Z^{\rm MF}_w$ and $Z_m = u Z^{\rm MF}_m$
hold up to $O(g^4)$ terms.

In the continuum we employ the $\ovl{\rm MS}$ scheme with naive dimensional 
regularization.  The one-loop self-energy in the continuum has the same form 
as \eqn{eqn:sigmaq2} with, however, the replacements,  
\begin{eqnarray}
\log (\lambda a)^2&\to&\log (\lambda/\mu)^2 , \\
\Sigma_1 & \to & \Sigma_1^{\ovl{\rm MS}}=1/2 , \\
\Sigma_2 & \to & \Sigma_2^{\ovl{\rm MS}}=-2 , \\
w_0 &\to& 0.
\end{eqnarray}

Let us define the quark wave function renormalization factor needed for 
converting the lattice field to the continuum field in the $\ovl{\rm MS}$ 
scheme by 
\begin{equation} 
q^{\ovl{\rm MS}}= (1-w_0^2)^{-1/2} Z_w^{-1/2} \sqrt{Z_2(\mu a)} q^{\rm lat}.
\end{equation}
To one-loop order we finally find that
\begin{equation}
Z_2 (\mu a) = 1+\frac{g^2}{16\pi^2}C_F\left[ 
-\log (\mu a)^2 + z_2 
\right] ,
\end{equation}
where 
\begin{equation}
z_2 = \Sigma_1^{\ovl{\rm MS}} -\Sigma_1 ,
\end{equation}
and $Z_w$ is given in \eqn{eqn:zw}.
The quark mass renormalization factor defined by
\begin{equation}
m^{\ovl{\rm MS}}(\mu )
= \left(1-w_0^2\right) Z_w Z_m (\mu a) m
\end{equation}
is given by
\begin{equation}
Z_m (\mu a) = 1+\frac{g^2}{16\pi^2}C_F\left[ 
-3 \log (\mu a)^2 +z_m \right] 
\end{equation}
with 
\begin{equation}
z_m = (\Sigma_2^{\ovl{\rm MS}} -\Sigma_2)
    - (\Sigma_1^{\ovl{\rm MS}} -\Sigma_1) .
\end{equation}

\reseteqnum
\section{Renormalization for quark bilinear operators}

\begin{figure}
\begin{center}\begin{picture}(200,100)(0,0)
\ArrowLine(180,10)(160,30)
\ArrowLine(160,30)(100,90)
\ArrowLine(100,90)(40,30)
\ArrowLine(40,30)(20,10)
\Gluon(40,30)(160,30){5}{5}
\Vertex(40,30){3}
\Vertex(160,30){3}
\Line(98,86)(102,94)
\Line(98,94)(102,86)
\Text(150,60)[l]{$l$}\LongArrow(145,65)(135,75)
\Text(65,75)[l]{$l$}\LongArrow(60,70)(50,60)
\Text(95,20)[l]{$l$}\LongArrow(100,20)(110,20)
\Text(95,100)[l]{$\ovl{q} \Gamma q$}
\Text(195,20)[l]{$p'=0$}\LongArrow(190,15)(180,25)
\Text(5,20)[r]{$p=0$}\LongArrow(20,25)(10,15)
\Text(35,35)[r]{$\bpsi_s \psi_s$}
\Text(165,35)[l]{$\bpsi_t \psi_t$}
\Text(20,0)[l]{$q$}
\Text(180,0)[l]{$\ovl{q}$}
\end{picture}\end{center}
\caption{One loop correction to the quark bilinear operator.}
\label{fig:bilinear}
\end{figure}
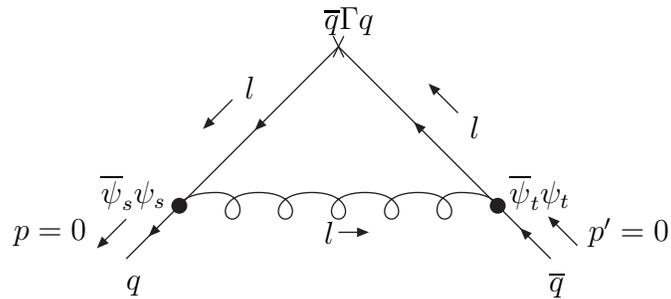

We consider quark bilinear operators in the following form
\begin{eqnarray}
{\cal O}_\Gamma(x) = \ovl{q}(x) \Gamma q(x)
,\qquad
\Gamma= 1, \gamma_5, \gamma_\mu, \gamma_\mu \gamma_5, \sigma_{\mu \nu}.
\label{eq:bilinear}
\end{eqnarray}
We calculate the one loop correction to the Green's function
$\vev{{\cal O}_\Gamma(x) q(y) \ovl{q}(z)}$ for massless quark
with external momenta $p=p^\prime=0$,
whose diagram is shown in Fig.~\ref{fig:bilinear}.
As in the previous section we notice that the external line is
essentially written in terms of the quark propagator times dumping
factors, \eqn{eqn:ext-line1} and \eqn{eqn:ext-line2}.
Making use of this fact 
the one loop full Green's function for small external momentum becomes
\begin{eqnarray}
\vev{(\ovl{q} \Gamma q) \cdot q \ovl{q}}_{\rm full}
=
\frac{\left(1-w_0^2\right) Z_w Z_2}{i\pslash} T_\Gamma \Gamma
\frac{\left(1-w_0^2\right) Z_w Z_2}{i\pslash'},
\end{eqnarray}
where $T_\Gamma$ is given by
\begin{eqnarray}
T_\Gamma =
1+\frac{g^2 C_F}{16 \pi^2}
\left( -\frac{h_2(\Gamma)}{4} \log \lambda^2 a^2 + V_\Gamma \right)
\end{eqnarray}
with a $\Gamma$ dependent constant 
$h_2(\Gamma) = 4(A), 4(V), 16(P), 16(S), 0(T)$.
The finite renormalization factor $V_\Gamma$ is given by
\begin{eqnarray}
V_\Gamma &=& \frac{h_2(\Gamma)}{4} \log \pi^2 + I_\Gamma\cdot 16\pi^2 ,
\\
I_\Gamma &=&
\int \frac{d^4 l}{(2\pi)^4}
\left[
r^2 \sin^2 \frac{l}{2} +r \frac{\sin^2 l}{e^{-\alpha}-W}
+ \frac{A_\Gamma}{\left(e^{-\alpha}-W\right)^2}
\right]
\frac{1}{\left(e^\alpha-w_0\right)^2} \frac{1}{4 \sin^2 \frac{l}{2}}
\nn\\&-&
\frac{h_2(\Gamma)}{4}
\int \frac{d^4 l}{(2\pi)^4} \frac{1}{(l^2)^2} \theta(\pi^2-l^2),
\end{eqnarray}
where
\begin{equation}
A_\Gamma =
\left\{
\begin{array}{ll}
\cos^2 \frac{l}{2} \sin^2 l
& \Gamma=S,\, P \\
\left( \cos^2 \frac{l}{2} -2 \cos^2 \frac{l_\alpha}{2} \right)
\left( \sin^2 l -2 \sin^2 l_\alpha \right)
& \Gamma=V_\alpha, \, A_\alpha \\
\left( \cos^2 \frac{l}{2}
 -2 \cos^2 \frac{l_\alpha}{2} -2 \cos^2 \frac{l_\beta}{2} \right)
\left( \sin^2 l -2 \sin^2 l_\alpha -2 \sin^2 l_\beta \right)
& \Gamma=T_{\alpha\beta} \\
\end{array}
\right. ,
\label{eqn:dirac}
\end{equation}
$\sin^2 l =\sum_\mu \sin^2 l_\mu$ and $\cos^2 l/2 = \sum_\mu \cos^2 l_\mu/2$.
The renormalization factor of quark bilinear operator is written as
\begin{eqnarray}
Z^q_\Gamma &=& (1-w_0^2)Z_w Z_2 T_\Gamma \equiv 
(1-w_0^2)Z_w Z_\Gamma^{\rm lat},
\\
Z_\Gamma^{\rm lat} &=&
1+\frac{g^2 C_F}{16 \pi^2}
\left[ \left(1-\frac{h_2(\Gamma)}{4} \right) \log \lambda^2 a^2
+ V_\Gamma +\Sigma_1 \right].
\end{eqnarray}
It is seen from \eqn{eqn:dirac} that $Z_A=Z_V$ and $Z_P=Z_S$,
which is expected to hold if the chiral symmetry exists and
is indeed so in the continuum QCD.

In the continuum, the on-shell vertex function to one-loop order 
is given in the $\ovl{\rm MS}$ scheme by 
\begin{equation}
T_\Gamma^{\ovl{\rm MS}} = 1 +\frac{g^2 C_F}{16\pi^2} \left(
\frac{h_2(\Gamma)}{4} \log(\mu/\lambda)^2 +V_\Gamma^{\ovl{\rm MS}}
\right)
\end{equation}
with
$V_\Gamma^{\ovl{\rm MS}} = -1/2(A), -1/2(V), 2(P), 2(S), 0(T)$
for the anti-commuting definition of $\gamma_5$.

Combining the above results and including self-energy corrections, 
the relation between the continuum operator in the $\ovl{\rm MS}$ scheme
and the lattice operator is given by 
\begin{equation}
O_\Gamma^{\ovl{\rm MS}}(\mu)=(1-w_0^2)^{-1}Z_w^{-1} Z_\Gamma(\mu a)
\left( \ovl{q} \Gamma q \right),
\label{eq:renop}
\end{equation}
where 
\begin{equation}
Z_\Gamma(\mu a) = 1+\frac{g^2 C_F}{16\pi^2}\left[ \left(\frac{h_2(\Gamma)}{4}
-1\right)\log(\mu a)^2 + z_\Gamma \right] ,
\end{equation}
and 
\begin{equation}
z_\Gamma = z_\Gamma^{\ovl{\rm MS}}-z_\Gamma^{\rm lat}
\end{equation}
with  
\begin{eqnarray}
z_\Gamma^{\ovl{\rm MS}} &=& V_\Gamma^{\ovl{\rm MS}}+1/2 
= 0(A), 0(V), 5/2(P), 5/2(S), 1/2(T),
 \\
z_\Gamma^{\rm lat} &=& V_\Gamma + \Sigma_1.
\end{eqnarray}

Here it should be mentioned that 
 $z_m = - z_S$ and consequently 
$Z_m(\mu a) = Z_S(\mu a)^{-1}$,
since $\sigma_m = I_S$ ( $\Sigma_2 = - V_S$ ).
To show this we use the relation
\begin{eqnarray}
\frac{(e^{-\alpha}-W)(e^\alpha -W)}{(e^{-\alpha}- W)^2}
&=&  \frac{ 1+W^2 - 2 W \cosh \alpha }{(e^{-\alpha}- W)^2}
= -\frac{\sin^2 l}{(e^{-\alpha}- W)^2} .
\end{eqnarray}
Note that the chiral Ward-Takahashi identities, if exact,
implies 
the three equalities $z_S = z_P = -z_m$ and $z_V = z_A$,
which is satisfied without fine tuning in the domain-wall QCD.

\section{Numerical results}

To calculate momentum integrals appeared in the previous two sections 
numerically,
two independent methods are employed, 
in order to check formulae and programs.
In the first method the momentum integration is performed 
by a mode sum for a periodic box of a size $L^4$ after transforming 
the momentum variable through $p_\mu=q_\mu-\sin q_\mu$.  
We employ the size $L=64$ for integrals.
In the second method the momentum integration is carried out by the 
Monte Carlo routine VEGAS, using 20 samples of 1000000 points each.  
We found that the both result agrees very well.

We first present basic quantities $\Sigma_1$, $\Sigma_3$,
$V_S ( =V_P )$, $V_V (=V_A)$, and $V_T$ in table~\ref{tab:sigma}
for several values of $M$.
Here $\Sigma_2$ is omitted since $\Sigma_2 = - V_S$ holds exactly.
The finite part of quark renormalization factors
$z_w$ and $z_2 $,
are given in table~\ref{tab:ZF}.
Since $z_m = - z_S$ we omit it here again.
We also give mean-field improved values where
$M$ should be replaced by $\wt{M}=M+4(u-1)$.
In table~\ref{tab:zgamma} we give our main result of this paper,
$z_{S,P}$, $z_{V,A}$ and $z_T$, for several values of $M$.
In these tables errors estimated from difference between two methods of
the numerical integration are in the (right) next to the last digit of 
each value.

\section{Conclusion}

In this paper we have calculated renormalization factors for quark bilinear
operators in domain-wall QCD at 1-loop of the perturbation theory.
We find that $z_S = z_P=-z_m$ and $z_V = z_A$, which suggest that
the chiral Ward-Takahashi identity holds exactly without fine tuning
at this order of the perturbation theory.
Our numerical values for $z_X$ with $X=2,m,S,P,
V,A,T$ are compared with those for the Wilson fermion action with and without
mean-field improvement and clover fermion action with $c_{SW}=1$, 
given in table~\ref{tab:others}. No data for clover fermion action with
mean-field improvement is given here since $c_{SW}$ itself is modified through 
the mean-field improvement.
Finite parts $z_X$ are rather large without mean-field improvement
in all three fermion formulations, but they becomes smaller after
the mean-field improvement is performed for the domain-wall fermion and
Wilson fermion.

Peculiar feature of the domain-wall QCD in its renormalization is an
appearance of the overlap factor $(1-w_0^2) Z_w$ for the physical
quark field. The problem here is that the 1-loop correction $z_w$ 
becomes very large unless $ \vert 1 - M \vert $ is small.
For example, at $M=1.7$, which is the value used in the quenched simulation 
at $\beta = 6.0$\cite{Blum-Soni}, from table~\ref{tab:ZF}
$z_w = 137.03 $ without mean-field improvement.
This 1-loop correction is huge, so we can not trust the perturbation theory.
If mean-field improvement is made with $u = P^{1/4}$ where $P$ is the
expectation value of the plaquette and is 0.59374 at $\beta=6.0$,
$z_w = 0.1893$ at $\wt{M} \simeq 1.2$. This value is reasonably small.
This example suggests that one have to employ the mean-field improvement for
perturbative renormalization factors in domain-wall QCD to convert
the lattice result to the one in the $\ovl{\rm MS}$ scheme.

\section*{Acknowledgements}

This work is supported in part by the Grants-in-Aid for
Scientific Research from the Ministry of Education, Science and Culture
(Nos. 2373, 2375). 
Y. T. and T. I are supported by Japan Society for Promotion of Science.

\newcommand{\J}[4]{{\it #1} {\bf #2} (19#3) #4}
\newcommand{\MPL}{Mod.~Phys.~Lett.}
\newcommand{\IJMP}{Int.~J.~Mod.~Phys.}
\newcommand{\NP}{Nucl.~Phys.}
\newcommand{\PL}{Phys.~Lett.}
\newcommand{\PR}{Phys.~Rev.}
\newcommand{\PRL}{Phys.~Rev.~Lett.}
\newcommand{\AP}{Ann.~Phys.}
\newcommand{\CMP}{Commun.~Math.~Phys.}
\newcommand{\PTP}{Prog. Theor. Phys.}
\newcommand{\Suppl}{Prog. Theor. Phys. Suppl.}


\begin{table}
\caption{Value of $\Sigma_1$, $\Sigma_3$ and $V_\Gamma$.}
\label{tab:sigma}
\begin{center}
\begin{tabular}{llllll}
$M$ & $\Sigma_1$  & $\Sigma_3$ & $V_{S,P}$ & $V_{V,A}$ & $V_T$ \\
\hline
0.050 & 13.25 & 51.22 & 3.30 & 4.833 & 5.344 \\
0.100 & 13.16 & 51.05 & 3.82 & 4.835 & 5.173 \\
0.150 & 13.08 & 50.89 & 4.24 & 4.836 & 5.035 \\
0.200 & 13.01 & 50.74 & 4.60 & 4.837 & 4.915 \\
0.250 & 12.94 & 50.61 & 4.93 & 4.838 & 4.807 \\
0.300 & 12.88 & 50.49 & 5.24 & 4.840 & 4.707 \\
0.350 & 12.83 & 50.37 & 5.53 & 4.84 & 4.613 \\
0.400 & 12.77 & 50.27 & 5.80 & 4.84 & 4.524 \\
0.450 & 12.72 & 50.17 & 6.06 & 4.85 & 4.439 \\
0.500 & 12.68 & 50.07 & 6.31 & 4.85 & 4.357 \\
0.550 & 12.63 & 49.98 & 6.56 & 4.85 & 4.277 \\
0.600 & 12.59 & 49.90 & 6.80 & 4.85 & 4.199 \\
0.650 & 12.56 & 49.83 & 7.04 & 4.85 & 4.122 \\
0.700 & 12.52 & 49.76 & 7.28 & 4.85 & 4.045 \\
0.750 & 12.49 & 49.69 & 7.52 & 4.86 & 3.970 \\
0.800 & 12.46 & 49.63 & 7.75 & 4.86 & 3.894 \\
0.850 & 12.44 & 49.58 & 7.99 & 4.86 & 3.819 \\
0.900 & 12.41 & 49.53 & 8.23 & 4.86 & 3.743 \\
0.950 & 12.39 & 49.49 & 8.47 & 4.87 & 3.666 \\
1.000 & 12.38 & 49.46 & 8.71 & 4.87 & 3.588 \\
1.050 & 12.36 & 49.43 & 8.96 & 4.87 & 3.509 \\
1.100 & 12.35 & 49.41 & 9.22 & 4.88 & 3.429 \\
1.150 & 12.34 & 49.39 & 9.48 & 4.88 & 3.346 \\
1.200 & 12.34 & 49.39 & 9.75 & 4.88 & 3.261 \\
1.250 & 12.34 & 49.39 & 10.02 & 4.89 & 3.173\\
1.300 & 12.35 & 49.40 & 10.31 & 4.89 & 3.082\\
1.350 & 12.36 & 49.42 & 10.62 & 4.89 & 2.987\\
1.400 & 12.37 & 49.45 & 10.93 & 4.90 & 2.888\\
1.450 & 12.39 & 49.49 & 11.27 & 4.90 & 2.783\\
1.500 & 12.42 & 49.54 & 11.62 & 4.91 & 2.673\\
1.550 & 12.45 & 49.61 & 12.00 & 4.92 & 2.555\\
1.600 & 12.49 & 49.69 & 12.40 & 4.92 & 2.428\\
1.650 & 12.54 & 49.80 & 12.84 & 4.93 & 2.292\\
1.700 & 12.60 & 49.92 & 13.31 & 4.94 & 2.142\\
1.750 & 12.68 & 50.07 & 13.84 & 4.94 & 1.978\\
1.800 & 12.77 & 50.25 & 14.42 & 4.95 & 1.793\\
1.850 & 12.87 & 50.46 & 15.09 & 4.96 & 1.582\\
1.900 & 13.00 & 50.72 & 15.88 & 4.97 & 1.334\\
1.950 & 13.15 & 51.03 & 16.84 & 4.98 & 1.027\\
\end{tabular}
\end{center}
\end{table}

\begin{table}
\caption{Value of $z_2$ and $z_w$. Mean-field improved values are also listed.}
\label{tab:ZF}
\begin{center}
\begin{tabular}{l|ll|ll}
 &\multicolumn{2}{c|}{}
    &\multicolumn{2}{c}{MF} \\
$M$ & $z_2$ & $z_w$ &$z_2$ & $z_w$ \\
\hline
0.050 & -12.75 & -998.23  & -0.52 & -44.70 \\
0.100 & -12.66 & -483.60  & -0.43 & -20.05 \\
0.150 & -12.58 & -311.75  & -0.35 & -11.99 \\
0.200 & -12.51 & -225.53  & -0.28 & -8.057  \\
0.250 & -12.44 & -173.52  & -0.21 & -5.759 \\
0.300 & -12.38 & -138.59  & -0.15 & -4.272 \\
0.350 & -12.33 & -113.39  & -0.09 & -3.244 \\
0.400 & -12.27 & -94.248  & -0.04 & -2.502 \\
0.450 & -12.22 & -79.114  & 0.01   & -1.946 \\
0.500 & -12.18 & -66.762  & 0.06 & -1.521 \\
0.550 & -12.13 & -56.408  & 0.10 & -1.188 \\
0.600 & -12.09 & -47.526  & 0.14 & -0.9253 \\
0.650 & -12.06 & -39.748  & 0.18 & -0.7147 \\
0.700 & -12.02 & -32.807  & 0.21 & -0.5446 \\
0.750 & -11.99 & -26.503  & 0.24 & -0.4062 \\
0.800 & -11.96 & -20.681  & 0.27 & -0.2929 \\
0.850 & -11.94 & -15.217  & 0.30 & -0.1995 \\
0.900 & -11.91 & -10.007  & 0.32 & -0.1218 \\
0.950 & -11.89 & -4.9617   & 0.34 & -0.05631 \\
1.000 & -11.88 & 0.0       & 0.36 & 0.0       \\
1.050 & -11.86 & 4.9553    & 0.37 & 0.04993 \\
1.100 & -11.85 & 9.9813    & 0.38 & 0.09620 \\
1.150 & -11.84 & 15.159   & 0.39 & 0.1416 \\
1.200 & -11.84 & 20.577   & 0.39 & 0.1893 \\
1.250 & -11.84 & 26.339   & 0.39 & 0.2428 \\
1.300 & -11.85 & 32.569   & 0.39 & 0.3066 \\
1.350 & -11.86 & 39.420   & 0.38 & 0.3863 \\
1.400 & -11.87 & 47.091   & 0.36 & 0.4894 \\
1.450 & -11.89 & 55.846   & 0.34 & 0.6264 \\
1.500 & -11.92 & 66.053   & 0.32 & 0.8119 \\
1.550 & -11.95 & 78.235   & 0.28 & 1.068 \\
1.600 & -11.99 & 93.173   & 0.24 & 1.427 \\
1.650 & -12.04 & 112.09  & 0.19 & 1.945 \\
1.700 & -12.10 & 137.03  & 0.13 & 2.711 \\
1.750 & -12.18 & 171.66  & 0.05 & 3.895 \\
1.800 & -12.27 & 223.31  & -0.03 & 5.840 \\
1.850 & -12.37 & 309.12  & -0.14 & 9.357 \\
1.900 & -12.50 & 480.47  & -0.26 & 16.91 \\
1.950 & -12.65 & 994.48  & -0.42 & 40.95 \\
\end{tabular}
\end{center}
\end{table}

\begin{table}
\caption{Value of $z_\Gamma$. Mean-field improved values are also listed.}
\label{tab:zgamma}
\begin{center}
\begin{tabular}{l|lll|lll}
  & \multicolumn{3}{c|}{} 
  &  \multicolumn{3}{c}{MF} \\
$M$ & $z_{S,P}=-z_m$ & $z_{V,A}$ & $z_T$ 
    & $z_{S,P}=-z_m$ & $z_{V,A}$ & $z_T$ \\
\hline
0.050 & -14.05 & -18.083 & -18.09 & -1.82 & -5.8497 & -5.86\\
0.100 & -14.48 & -17.996 & -17.83 & -2.25 & -5.7627 & -5.60\\
0.150 & -14.82 & -17.918 & -17.62 & -2.59 & -5.6853 & -5.38\\
0.200 & -15.11 & -17.848 & -17.43 & -2.88 & -5.6149 & -5.19\\
0.250 & -15.38 & -17.783 & -17.25 & -3.15 & -5.5503 & -5.02\\
0.300 & -15.62 & -17.723 & -17.09 & -3.39 & -5.4906 & -4.86\\
0.350 & -15.85 & -17.668 & -16.94 & -3.62 & -5.4352 & -4.71\\
0.400 & -16.07 & -17.617 & -16.80 & -3.84 & -5.3838 & -4.57\\
0.450 & -16.28 & -17.569 & -16.66 & -4.05 & -5.3359 & -4.43\\
0.500 & -16.49 & -17.524 & -16.53 & -4.26 & -5.2913 & -4.30\\
0.550 & -16.70 & -17.483 & -16.41 & -4.46 & -5.2500 & -4.18\\
0.600 & -16.90 & -17.444 & -16.29 & -4.67 & -5.2117 & -4.06\\
0.650 & -17.10 & -17.409 & -16.18 & -4.87 & -5.1764 & -3.95\\
0.700 & -17.30 & -17.377 & -16.07 & -5.07 & -5.1440 & -3.84\\
0.750 & -17.51 & -17.347 & -15.96 & -5.27 & -5.1145 & -3.73\\
0.800 & -17.71 & -17.321 & -15.86 & -5.48 & -5.0879 & -3.62\\
0.850 & -17.92 & -17.297 & -15.75 & -5.69 & -5.0643 & -3.52\\
0.900 & -18.14 & -17.277 & -15.66 & -5.91 & -5.0437 & -3.42\\
0.950 & -18.36 & -17.259 & -15.56 & -6.13 & -5.0262 & -3.33\\
1.000 & -18.59 & -17.245 & -15.46 & -6.35 & -5.0119 & -3.23\\
1.050 & -18.82 & -17.234 & -15.37 & -6.59 & -5.0012 & -3.14\\
1.100 & -19.07 & -17.226 & -15.28 & -6.84 & -4.9935 & -3.05\\
1.150 & -19.32 & -17.223 & -15.19 & -7.09 & -4.9898 & -2.96\\
1.200 & -19.59 & -17.223 & -15.10 & -7.35 & -4.9900 & -2.87\\
1.250 & -19.87 & -17.227 & -15.01 & -7.63 & -4.9944 & -2.78\\
1.300 & -20.16 & -17.236 & -14.93 & -7.93 & -5.0034 & -2.70\\
1.350 & -20.47 & -17.250 & -14.84 & -8.24 & -5.0174 & -2.61\\
1.400 & -20.80 & -17.270 & -14.76 & -8.57 & -5.0368 & -2.53\\
1.450 & -21.16 & -17.295 & -14.67 & -8.92 & -5.0621 & -2.44\\
1.500 & -21.54 & -17.327 & -14.59 & -9.30 & -5.0941 & -2.36\\
1.550 & -21.95 & -17.366 & -14.51 & -9.71 & -5.1335 & -2.27\\
1.600 & -22.39 & -17.414 & -14.42 & -10.16 & -5.181 & -2.19\\
1.650 & -22.88 & -17.471 & -14.34 & -10.65 & -5.238 & -2.10\\
1.700 & -23.42 & -17.539 & -14.25 & -11.18 & -5.307 & -2.01\\
1.750 & -24.01 & -17.620 & -14.16 & -11.78 & -5.388 & -1.92\\
1.800 & -24.69 & -17.716 & -14.06 & -12.46 & -5.484 & -1.83\\
1.850 & -25.46 & -17.831 & -13.95 & -13.23 & -5.598 & -1.72\\
1.900 & -26.37 & -17.968 & -13.83 & -14.14 & -5.735 & -1.60\\
1.950 & -27.49 & -18.135 & -13.68 & -15.26 & -5.902 & -1.45\\
\end{tabular}
\end{center}
\end{table}

\begin{table}
\caption{Value of $z_X$ for Wilson and clover fermion actions. }
\label{tab:others}
\begin{center}
\begin{tabular}{llll}
  & Wilson & Clover($c_{SW}=1$) & Wilson(MF) \\
\hline
$ z_2$ & -12.852 & -9.206 & -0.619 \\
$ z_m$ &  12.953 & 19.311 & 0.719  \\
$ z_S$ & -12.953  & 12.661 & -0.72  \\
$ z_P$ & -22.596  & 9.602  & -10.363  \\
$ z_V$ & -20.618  & 16.657 & -8.385 \\
$ z_A$ & -15.797  & 18.193 & -3.564 \\
$ z_T$ & -17.018  & 20.824 & -4.785 \\
\end{tabular}
\end{center}
\end{table}

\begin{figure}
\begin{center}\begin{picture}(200,100)(0,0)
\Text(-15,50)[r]{$V_{1\mu}^a(k,p)$}
\ArrowLine(100,70)(10,90)
\Text(10,90)[r]{$\bar\psi(k)$}
\Text(25,90)[lb]{$\rightarrow k$}
\ArrowLine(190,90)(100,70)
\Text(193,90)[l]{$\psi(p)$}
\Text(175,90)[rb]{$p \leftarrow$}
\Gluon(100,10)(100,70){5}{5}
\Text(115,45)[l]{$l$}\LongArrow(115,25)(115,35)
\Text(85,45)[r]{$A^a_\mu(l)$}
\Vertex(100,70){3}
\end{picture}\end{center}
\begin{center}\begin{picture}(200,100)(0,0)
\Text(-15,50)[r]{$V_{2\mu\nu}^{ab}(k,p)$}
\ArrowLine(100,70)(10,90)
\Text(10,90)[r]{$\bar\psi(k)$}
\Text(25,90)[lb]{$\rightarrow k$}
\ArrowLine(190,90)(100,70)
\Text(193,90)[l]{$\psi(p)$}
\Text(175,90)[rb]{$p \leftarrow$}
\Gluon(50,10)(100,70){5}{5}
\Text(65,55)[l]{$l_1$}\LongArrow(40,30)(60,50)
\Text(45,15)[r]{$A^{a}_\mu(l_1)$}
\Gluon(150,10)(100,70){5}{5}
\Text(135,55)[r]{$l_2$}\LongArrow(160,30)(140,50)
\Text(155,15)[l]{$A^{b}_\nu(l_2)$}
\Vertex(100,70){3}
\end{picture}
\end{center}
\caption{Momentum assignment of the fermion-gluon vertices.}
\label{fig:vertex}
\end{figure}
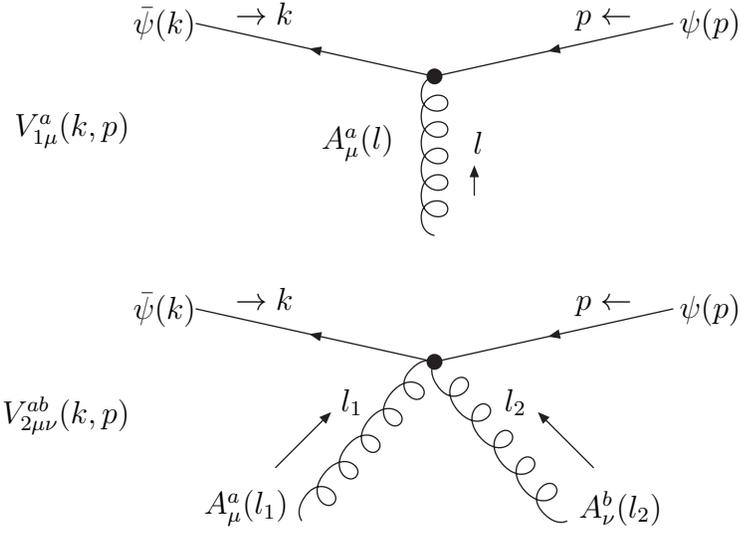

\begin{figure}
\begin{center}\begin{picture}(200,100)(0,0)
\GlueArc(100,50)(30,-90,270){-5}{12}
\Vertex(100,20){3}
\ArrowLine(100,20)(10,20)
\ArrowLine(180,20)(100,20)
\Text(160,30)[b]{$\leftarrow p$}
\Text(40,30)[b]{$\leftarrow p$}
\Text(150,60)[r]{$l$}\LongArrow(143,65)(138,75)
\Text(5,20)[r]{$q(-p)$}
\Text(185,20)[l]{$\ovl{q}(p)$}
\Text(100,5)[c]{$\bpsi_s \psi_s$}
\end{picture}\end{center}
\begin{center}\begin{picture}(200,100)(0,0)
\GlueArc(100,20)(50,0,180){-5}{12}
\Vertex(50,20){3}
\Vertex(150,20){3}
\ArrowLine(50,20)(10,20)
\ArrowLine(150,20)(50,20)
\ArrowLine(180,20)(150,20)
\Text(180,30)[b]{$\leftarrow p$}
\Text(100,5)[b]{$\leftarrow l$}
\Text(20,30)[b]{$\leftarrow p$}
\Text(155,60)[l]{$(p-l)$}\LongArrow(147,65)(137,75)
\Text(5,20)[r]{$q(-p)$}
\Text(190,20)[l]{$\ovl{q}(p)$}
\Text( 50,10)[c]{$\bpsi_s \psi_s$}
\Text(150,10)[c]{$\bpsi_t \psi_t$}
\end{picture}\end{center}
\caption{One loop correction to the quark propagator.}
\label{fig:propagator}
\end{figure}
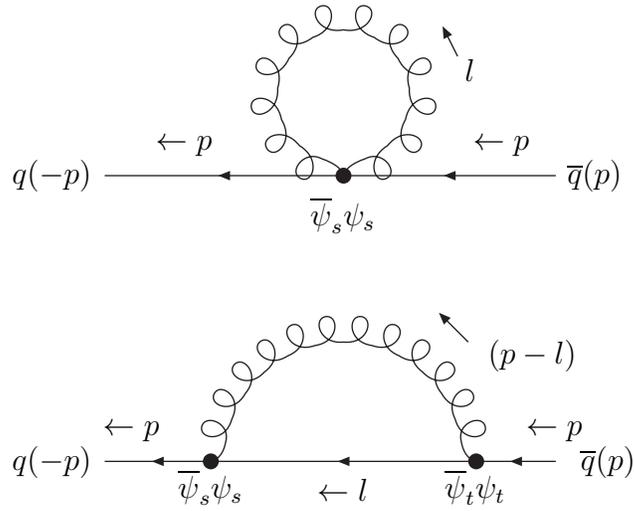

\end{document}